\documentclass[letterpaper, 10 pt, conference]{ieeeconf}  

\IEEEoverridecommandlockouts                              

\usepackage{graphics} 
\usepackage{amsmath} 
\usepackage{amssymb}  
\usepackage{color}
\usepackage{epstopdf}

\usepackage[demo]{graphicx}

{}
{}
{}
\newtheorem{theorem}{Theorem}{}
\newtheorem{remark}{Remark}{}
\newtheorem{lemma}{Lemma}{}

\newcommand{\overbar}[1]{\mkern 1.5mu\overline{\mkern-1.5mu#1\mkern-1.5mu}\mkern 1.5mu}

\title{\LARGE \bf
Filtering Approaches for Dealing with Noise in Anomaly Detection
}

\author{Navid Hashemi$^{1}$, Eduardo Verdugo German$^{2}$, Jonatan Pena Ramirez$^{2}$, and Justin Ruths$^{1}$
\thanks{*This work is supported by the ConTex grant award 2018-56A and 2018-56B .}
\thanks{$^{1}$These authors are with the Departments of Mechanical and Systems Engineering at the University of Texas at Dallas, Richardson, Texas, USA 
        {\tt\small navid.hashemi@, jruths@utdallas.edu}}%
\thanks{$^{2}$These authors are with the Department of Electronics and Telecommunications at the Scientific Research and Advanced Studies Center of Ensenada, Mexico (CICESE)
        {\tt\small jverdugo@cicese.edu.mx, jpena@cicese.mx}}%
}

\begin{document}

\maketitle
\thispagestyle{empty}
\pagestyle{empty}

\begin{abstract}

The leading workhorse of anomaly (and attack) detection in the literature has been residual-based detectors, where the residual is the discrepancy between the observed output provided by the sensors (inclusive of any tampering along the way) and the estimated output provided by an observer.  These techniques calculate some statistic of the residual and apply a threshold to determine whether or not to raise an alarm. To date, these methods have not leveraged the frequency content of the residual signal in making the detection problem easier, specifically dealing with the case of (e.g., measurement) noise.  Here we demonstrate some opportunities to combine filtering to enhance the performance of residual-based detectors. We also demonstrate how filtering can provide a compelling alternative to residual-based methods when paired with a robust observer.  In this process, we consider the class of attacks that are stealthy, or undetectable, by such filtered detection methods and the impact they can have on the system.

\end{abstract}

\section{INTRODUCTION}

In anomaly detection, we seek to differentiate normal behavior from anomalous behavior - essentially anything that deviates away from the nominal model. This task is easy if our confidence is high in the nominal model but as uncertainty rises, distinguishing normal from anomalous becomes more challenging. From a control theory perspective, one of the most fundamental sources of uncertainty is measurement noise and in this work we present an intuitive, but to-date-unexplored, idea of combining traditional model-based detection schemes with low-pass filtering to reduce the impact of noise on our ability to detect anomalies in control systems. Beyond demonstrating that this approach enables better detection, the novelty lies in characterizing this performance boost analytically, in characterizing how this effects the impact an attacker can have on the system, and in understanding what new kind of attacks are possible when filtering is involved.

There is a deep literature from Fault Detection that leverages model-based detectors to identify the occurrence of faults \cite{Patton_1}. More recently these tools have been reused in identifying the presence of attacks in control systems \cite{Cardenas}\nocite{mo2014detecting,Mo_3,Guo2016,milovsevic2018quantifying,umsonst2018anomaly,RuthsACC2018_Hidden,RuthsACC2018_Windowed,RuthsACC2018_Comparison}-\cite{RuthsACC2019_nongaussian}. Attacks, in particular their strategic and exploitative nature, offer new challenges to both control theory and anomaly detection. While many research groups have rallied behind the banner of analyzing attacks in control systems and the challenges they raise, in what follows we describe the main representative contributions of these groups that relate to characterizing the tuning and performance of model-based detectors. 

While all papers \cite{Cardenas}-\cite{RuthsACC2019_nongaussian} investigate the use of model-based detectors, and \cite{Cardenas}-\cite{RuthsACC2018_Comparison} consider various types of attacks, it has only been recently that the impact of these attacks has been evaluated \cite{milovsevic2018quantifying}-\cite{RuthsACC2018_Comparison}. Of these \cite{Mo_3,RuthsACC2018_Hidden,RuthsACC2018_Comparison} use the reachable set to characterize attack impact and \cite{milovsevic2018quantifying,umsonst2018anomaly,RuthsACC2018_Windowed} use various norms of system state or the estimation error covariance \cite{Guo2016}. To a large extent this work on attacker impact has been facilitated by analytic methods to tune model-based detectors, as provided in \cite{Mo_3,umsonst2018anomaly,RuthsACC2018_Hidden,RuthsACC2018_Windowed,RuthsACC2018_Comparison,RuthsACC2019_nongaussian} and for work to define various types of worst-case attacks \cite{mo2014detecting}, \cite{Guo2016}-\cite{RuthsACC2018_Comparison}. In all of these cases Gaussian noise(s) are assumed on the measurements of the system (and possibly the system state), except in \cite{RuthsACC2019_nongaussian} which tunes a detector for generalized noise distributions.

While much has been done with model-based detectors, no study has integrated a low-pass filter to attenuate the effect of the noise on the attacker's capabilities to change the system state. The recent work to to tune classical detectors to desired levels of performance (desired false alarm rate) and to characterize the attacker's impact on the system state (given the attacker's desire to remain stealthy to classical detectors) positions us well to now add another layer, i.e., filtering, to the detection scheme. In Section \ref{sec:residual_detection}, we characterize the improvement gained by adding the filter (in terms of sensitivity to attacks and reduction in attacker capabilities) traded off with the new attack definitions that are stealthy to the introduced \textit{low-pass chi-squared detector}.

This notion of filtering is not exclusive to retrofit conventional residual-based detectors, but generalizes in compelling ways to other types of model-based observers. In particular, in Section \ref{sec:discontinuous_observer}, we review how the discontinuous observer, an estimator that uses a sliding mode to yield finite-time convergence, produces a discontinuous term that can be filtered to produce an estimate of the disturbance that enters a system \cite{robustobs}. In this case, we reframe the disturbance estimation problem as a method to approximate the anomaly (e.g., failure and/or attack), which provides an attractive alternative to residual-based detectors. In this case the disturbance estimate not only provides a way to detect the presence of an anomaly, but also find the form of the anomaly so that advanced mitigation strategies can be employed - such as using the estimated anomaly as feedback to avoid the estimate to deviate away from the true state value. 

In Section \ref{sec:system} we introduce the system and our notation as well as the attack context; in Section \ref{sec:example} we provide an example that combines and compares both of these tools.

\section{SYSTEM \& CONTEXT} \label{sec:system}
We consider a continuous, linear, time-invariant control system
 \begin{equation} \label{eq:dLTIsystem}
 \left\{\begin{aligned}
 	\dot{x}(t) &= Ax(t) + Bu(t),\\
 	y(t) &= Cx(t) + \eta(t), \\
 \end{aligned}\right.
 \end{equation}
in which $x(t)\in\mathbb{R}^n$ is the system state, $u(t)\in\mathbb{R}^m$ is the input, A$\in\mathbb{R}^{n\times n}$ is the system matrix, $B\in\mathbb{R}^{n\times m}$ is the input matrix, $y\in\mathbb{R}^p$ is the system output, and $\eta(t) \in \mathbb{R}^p$ is zero mean Gaussian additive measurement noise with covariance $R$, i.e., $\eta(t) \sim \mathcal{N}(0,R)$. The pair $(A,C)$ is detectable and the pair $(A,B)$ is stablizable.

We use the output $y$ and an estimator to produce an estimate of the state, $\hat{x}(t)$. In this work we consider two different choices of estimators. We use this estimate for to construct the feedback law,
\begin{equation}
    u(t)=K\hat{x}(t).
\end{equation}
The process measurements are potentially corrupted by an attacker, who is able to read and arbitrarily change the sensor measurements. We model this as an additive attack $\delta(t) \in \mathbb{R}^p$ such that $y(t)$ is changed to $\bar{y}(t)$,
 \begin{equation}\label{outputb}
 	\bar{y}(t) = y(t) + \delta(t) =  Cx(t) + \eta(t) + \delta(t).
 \end{equation}

\subsection{Attacker Capabilities}
In this work we assume that the attacker has access to all system information including, e.g., dynamics, states, estimator states, and detector parameters. In particular, the attacker can view (disclosure) and edit (disruption) sensor measurements.  

\section{RESIDUAL-BASED ANOMALY DETECTION} \label{sec:residual_detection}

The residual quantifies the difference between what we receive from the sensor, $\bar{y}(t)$ and what we expect based on an estimator, $\hat{y}(t)=C\hat{x}(t)$, 
\begin{equation}
	r(t) = \bar{y}(t) - C\hat{x}(t),
\end{equation}
Anomaly detectors use this quantity to make real-time choices in the state of the control system they monitor.

To build an estimate of the system state we use a Luenberger observer with observer gain $L \in \mathbb{R}^{n \times p}$,
\begin{equation}
\begin{aligned}
	\dot{\hat{x}}(t) =& A\hat{x}(t) + Bu(t) + L(\bar{y}(t) - C\hat{x}(t)).
\end{aligned}
\end{equation}
Armed with this estimator, we can express the difference between the state and expected state as the estimation error $e(t)=x(t)-\hat{x}(t)$, which leads to the following coupled closed-loop equations,
\begin{equation} \label{eq:esterror}
\left\{ \begin{aligned}
    \dot{x}(t) &= \big( A + BK \big) x(t) - BKe(t), \\
	\dot{e}(t) &= \big( A - LC \big) e(t) - L\eta(t) - L \delta(t), \\
	r(t) &= Ce(t) + \eta(t) + \delta(t).
\end{aligned}\right.
\end{equation}
The steady state Kalman filter is the standard choice of estimator used when employing residual based detectors. This allows us to calculate the steady state covariance of the estimation error from the stochastic dynamics of $e(t)$ using the following Riccati equation,
\begin{equation}
\dot{P}=0=(A-LC) P + P(A-LC)'+LR L'.
\end{equation}

Because the detection process is inherently a sampled approach we discretize \eqref{eq:esterror} with uniform step size $\tau$ such that $\xi(k\tau)=\xi_k$,
\begin{equation} \label{eq:discrete}
\left\{ \begin{aligned}
    x_{k+1} &= F x_k + G e_k, \\
	e_{k+1} &= H e_k - L_d(\eta_k + \delta_k), \\
	r_k &= Ce_k + \eta_k + \delta_k.
\end{aligned}\right.
\end{equation}
Residual-based detection relies on quantifying the distribution of the residual under nominal operation, i.e., without attacks/anomalies. These statistics form a one-sided hypothesis test that is either accepted (no anomalies) or rejected (alarm raised). The discrete covariance of $\eta_k$ is $\Sigma_\eta = \tau R$ and the discrete covariance of the estimation error is $\Sigma_e = \tau P$.
In the absence of attack the distribution of the residual follows a zero-mean Gaussian distribution with the covariance \cite{Carlos_Justin2},
\begin{equation} \label{eq:residual covariance}
	\Sigma_r = E[r_kr'_k] = C\Sigma_e C'+\Sigma_\eta.
\end{equation}
In the following subsections we will review the conventional chi-squared detector and subsequently extend this detector to enhance its performance.

\subsection{Chi-Squared Detector}
The role of the detector is to create a non-negative scalar-valued random variable from the residual that can be easily compared with a threshold. One of the most widely used approaches to do this is given by the chi-squared detector which introduces a distance measure
\begin{equation} \label{eq:distancemeasure}
	z(t)= r'(t)\Sigma_r^{-1}r(t).
\end{equation}
Since the residual is normally distributed, the distance measure, as a sum of squared Gaussian random variables, is chi-squared distributed (with $p$ degrees of freedom, where $p$ is the number of the sensors). By normalizing the residual covariance, we eliminate system dependence. The detection procedure for the chi-squared detector is summarized as follows. For given a threshold $\alpha \in \mathbb{R}_{>0}$ and the distance measure in \eqref{eq:distancemeasure}
\begin{equation} \label{detector}
\left\{\begin{aligned}
	z(t) \leq \alpha &\quad\longrightarrow\quad \text{no alarm}, \\
	z(t) > \alpha &\quad\longrightarrow\quad \text{alarm}.
\end{aligned}\right.
\end{equation}
Since the chi-squared distribution has $p$ degrees of freedom the distance measure has mean $E[z(t)]=p$ and covariance $E[z^2]=2p$ during nominal (no anomaly) operation. In the presence of attacks/anomalies the distance measure will, in general, not fall according to a chi-squared distribution with mean value $p$ and covariance $2p$. As indicated in \eqref{detector}, the detector is required to make a decision to raise or not raise an alarm at every time instant. Therefore, it is impractical to build a sample distribution of ``current behavior'' to decide if the distance measure was still distributed in chi-squared fashion - this would take too long. Instead, the alarms, and more specifically, the rate of alarms can be used as a metric for how deviated the system is from nominal behavior. 

Here \textit{alarm rate} is defined as the rate of generation of alarms by the detector, which empirically is the fraction of time instants in which an alarm was raised. We expect alarms to be raised regularly, even under normal operation, because of the infinite support of the measurement noise. We can predict the false alarm rate $\mathcal{A}$ - the alarm rate under normal operation - from the distance measure distribution, $\textbf{Pr}(z > \alpha)= \mathcal{A}$. More importantly, we can set the false alarm rate to a desired value $\mathcal{A}^*$ by selecting the threshold $\alpha$ appropriately.
\begin{lemma} \cite{Carlos_Justin2}. \label{lem:chisquared_tuning}
Consider a stochastic LTI system under normal operation and a chi-squared detector, with threshold $\alpha \in \mathbb{R}_{>0}$, $r_k \sim \mathcal{N}(0,\Sigma)$. Let $\alpha = \alpha^{*}:=2P^{-1}(1-\mathcal A^{*}, \frac{p}{2})$, where $P^{-1}(\cdot,\cdot)$ denotes the inverse regularized lower incomplete gamma function, then ${\mathcal A}= {\mathcal A}^{*}$. 
\end{lemma} 
If the attacker wants to remain stealthy (undetected) to the detector, the threshold of the chi-squared detector puts a limit on the distance measure. It, therefore, limits the capabilities of stealthy attacks to change the system behavior. We derive worst-case stealthy attacks by assuming a powerful, knowledgeable attacker that knows the noisy measurement $Cx+\eta$ and also knows the state estimate $\hat{x}$, the attack can compensate for the terms in the residual
\begin{equation}\label{eq:deltabar}
	\delta_k = - \bar{y}_k+C\hat{x}_k+\Sigma_r^{\frac{1}{2}}\bar{\delta}_k,
\end{equation}
such that
\begin{equation}
    z_k=\bar{\delta}'_k\bar{\delta}_k.
\end{equation}
By increasing the norm of $\bar{\delta}_k$ the attacker can increase the impact of the attack up to the point where the attack is identified by the detector; the ``direction'' of the vector $\bar{\delta}_k$ indicates the extent to which the attack impacts each sensor.

In \textit{\underline{zero alarm attacks}} the attacker takes the perspective that raising no alarms is the way to avoid detection, essentially keeping the distance measure below the threshold,
\begin{equation}
    z_k=\bar{\delta}'_k\bar{\delta}_k \leq \alpha,
\end{equation}
and hence $\| \bar{\delta}_k \| \leq \sqrt{\alpha}$.

In \textit{\underline{hidden attacks}}, the attacker realizes that under normal operation alarms are generated at the rate of $\mathcal{A}^*$, hence, it is reasonable to generate an attack sequence that mimics the false alarm rate,
\begin{equation} \label{eq:hiddendef}
	\textbf{Pr}(z_k > \alpha) = \textbf{Pr}\big(\|\bar\delta_k \| > \sqrt{\alpha}\big) = \mathcal{A}^*.
\end{equation}
Hidden attacks are inherently more potent than zero-alarm attacks, especially when keeping in mind that the norm of the attack can be arbitrarily large at the time instants in which alarms are raised. 

For more detail on these attacks and ways in which to evaluate the impact of these attacks through state deviation or the induced reachable set see \cite{RuthsACC2018_Windowed,RuthsACC2018_Comparison}.

\subsection{Filtered Chi-Squared Detector}

In this section we introduce a modification to the chi-squared detector to leverage apriori knowledge that the noise is typically composed of substantially higher frequencies than the disturbances or many types of anomalies. By filtering the residual signal, we aim to reduce the covariance of the statistic that forms the distance measure so that attacks are more readily apparent. Exploiting frequency domain information has not been combined with attack detection and here we introduce the approach and characterize its performance.

We filter the residual element-wise using a bank of $p$ identical Butterworth filters (of second order) such that $\xi_i(t)=[\xi_{1i}(t),\xi_{2i}(t)]'$ and
\begin{equation} \label{eq:filter_dynamics}
    \left\{\begin{aligned}
    \dot\xi_i(t) &= \Phi \xi_i(t)+ \Psi r_i(t),\\
    \rho_i(t) &= [1\ 0]\xi_i(t)
    \end{aligned}\right.
\end{equation}
for $i=1,\dots,p$ and where
\begin{equation} \label{eq:filter_matrices}
        \Phi=\begin{bmatrix}0&1\\-\omega_c^2&-\sqrt{2}\omega_c\end{bmatrix},\ \  \Psi=\begin{bmatrix}0\\\omega_c^2\end{bmatrix}.
\end{equation}
and we use the filtered output $\rho(t)$ to construct a new distance measure. To characterize the distribution of this distance measure, first characterize the  covariance of $\rho_i(t)$.
\begin{theorem}
Given a residual signal $r_k\sim \mathcal{N}(0,\Sigma_r)$ filtered through the Butterworth filter in \eqref{eq:filter_dynamics}-\eqref{eq:filter_matrices} with bandwidth $\omega_c$, the discrete-time filter with sampling rate $\tau$ output $\rho_k$ is zero mean Gaussian with covariance,
\begin{equation}\label{eq:filtered_covariance}
    \Sigma_\rho = \frac{\tau \omega_c}{2\sqrt{2}}\Sigma_r.
\end{equation}
\end{theorem}
\vspace{2mm}

\begin{proof}
We first discretize the filter with time step $\tau$
\begin{equation} \label{eq:filter_dynamics_discrete}
    \left\{\begin{aligned}
    \xi_k^{(i)} &= \Phi_d \xi^{(i)}_k+ \Psi_d r^{(i)}_k,\\
    \rho_k^{(i)} &= [1\ 0]\xi^{(i)}_k
    \end{aligned}\right.
\end{equation}
where $\zeta^{(i)}_k = \zeta_i(k\tau)$ and for small enough $\tau$, $\Phi_d=I+\Phi\tau$ and $\Psi_d=\Psi\tau$. To find the covariance of the filtered residual vector $\rho_k$ we collect all filters together,
\begin{equation}\label{eq:combinedfilter}
\dot{\xi}_k = \widetilde{\Phi} \xi_k + \widetilde{\Psi} r_k,
\end{equation}
with $\widetilde{\Phi}=\Phi_d\otimes I_p$ and $\widetilde{\Psi}=\Psi_d\otimes I_p$ and where $\xi_k \in \mathbb{R}^{2p}$, $\widetilde{\Phi} \in \mathbb{R}^{2p\times2p}$ and $\widetilde{\Psi} \in \mathbb{R}^{2p\times p}$. We can define the covariance of $\xi_k$ as $\mathcal{P}=[\mathcal{P}_{i,j}]$, $i=1,\dots,p$ and $j=1,\dots,p$, where $\mathcal{P}_{i,j} \in \mathbb{R}^{2 \times 2}$. This total covariance is supplied by the discrete-time Riccati equation (in steady state), 
\begin{equation}\label{eq:combinedfilter}
    0 = \widetilde{\Phi}\mathcal{P}\widetilde{\Phi}'-\mathcal{P} + \widetilde{\Psi}\Sigma_r\widetilde{\Psi}',
\end{equation}
and following some algebra manipulation reveals that we can split this into Riccati equations on each subblock
\begin{equation}\label{eq:splittedfilter}
\Phi_d \mathcal{P}_{i,j} \Phi_d' - \mathcal{P}_{i,j} + \Psi_d[\Sigma_r]_{i,j} \Psi_d'=0.
\end{equation}
From \eqref{eq:splittedfilter} we see that $\mathcal{P}_{i,j}=\mathcal{P}_{j,i}$ because $[\Sigma_r]_{i,j}=[\Sigma_r]_{j,i}$ and in addition based on \eqref{eq:combinedfilter}, we know that $\mathcal{P}_{i,j}=\mathcal{P}_{j,i}'$, therefore matrix $\mathcal{P}_{i,j}$ is a symmetric square matrix. The covariance of $\rho(t)$ can be computed element-wise using the structure of $\Phi_d$ and $\Psi_d$ leaving,
\begin{align}
    [\Sigma_\rho]&_{i,j} = [\mathcal{P}_{i,j}]_{11}\\ &= -\tau\omega_c {\tiny \frac{(\tau\omega_c)^2-\sqrt{2}\tau\omega_c+2}{(\tau\omega_c)^3-3\sqrt{2}(\tau\omega_c)^2+8\tau\omega_c-4\sqrt{2}}} \Sigma_r \nonumber
\end{align}
If the sampling time is taken small enough the higher order terms vanish leaving
\begin{equation}
    [\Sigma_{\rho}]_{i,j}=\frac{\tau\omega_c}{2\sqrt{2}} [\Sigma_r]_{i,j},
\end{equation}
which leads directly to \eqref{eq:filtered_covariance}.
\end{proof}

\begin{remark}
Note that $\tau$ is the sampling time of the system, which highlights that the filtered covariance is expected to be significantly smaller than the covariance of the original residual. By reducing the covariance of the nominal behavior using this filtering technique the aim is to make it easier to distinguish attacks and anomalies (especially those with low frequency components).
\end{remark}

Using this covariance, we can create a new normalized distance measure
\begin{equation}
    z_k=\rho'_k \Sigma_\rho^{-1} \rho_k = \frac{2\sqrt{2}}{\tau\omega_c} \rho'_k \Sigma_r^{-1} \rho_k.
\end{equation}
Because $\rho_k$ is a Gaussian random variable, $z(t)$ is also chi-squared. In fact, because of the normalization by the covariance, this functions exactly like the conventional chi-squared detector (including using Lemma \ref{lem:chisquared_tuning} to select the threshold). The major difference lies in the scale of the normalizing covariance; here being smaller provides an advantage to identify smaller attacks. To execute these attacks, attackers will also need to know the cut-off frequency of the filter. Analogous to \eqref{eq:deltabar}, we define
\begin{equation}
    \delta_k=-y_k+C\hat{x}_k+\left(\frac{\tau\omega_c}{2\sqrt{2}}\Sigma_r\right)^{\frac{1}{2}}\ \bar{\delta}_k,
\end{equation}
where $\bar{\delta}_k$ is an independent random variable that attackers may select to define different classes of stealthy attacks.

As before the \textit{\underline{zero alarm attack}} is generated in a way so that no alarm is generated by the detector. The precise characterization of the sequence $\bar{\delta}_k$ is a topic for future work, however, intuitively the fact that the attack signal is also filtered by the low-pass filter, it follows that the frequency content (as quantified by a Fourier Transform) and not just the amplitude (norm) of the attack vector plays a role in quantifying zero alarm attacks. To start, note that zero alarm attacks of the conventional chi-squared detector will also be zero alarm attacks of the filtered chi-squared detector. On top of this, there is an opportunity to inject high frequency content into the attack, which would be attenuated by the action of filter. If the filtered chi-squared used an \textit{ideal} low pass filter, it would be possible to inject any signal that had frequency content higher than the cut-off frequency of the filter. The second order filter requires some adjustment to balance the small contributions from the stopband frequencies with that of the passband frequencies.  

Similar to the unfiltered case, we employ the zero alarm attack constraint in a probabilistic fashion for the corresponding \textit{hidden attack}. In this case $\bar{\delta}_k$ is selected such that the Fourier frequency content is now bounded in probability. Again the full characterization of these attacks are for future work.

\section{DISCONTINUOUS OBSERVER} \label{sec:discontinuous_observer}
In this section, we present an anomaly detector for \textit{smooth} attacks $\delta(t)$, based on a robust observer and a low pass filter, which is applicable  for a particular class of second order systems, i.e., systems of the form (\ref{eq:dLTIsystem}) with $x=[x_{1}\ x_{2}]^{T}$, $A\in\mathbb{R}^{2\times2}$, $B\in\mathbb{R}^{2}$, and $C\in\mathbb{R}^{1\times2}$. Note, however, that any physical plant that can be partitioned into a collection of second order systems can be captured by an extended version of the presented methods. 

The observer is given by  \cite{robustobs}
\begin{align}
\dot{\hat{x}}&=A\hat{x}+Bu+\Gamma e+Bc_{3}\hbox{sign}(\bar{y}-\hat{y}),\label{obsi1}\\
\hat{y}&=C\hat{x},\label{obsi2}
\end{align}
where  $\hat{x}=[\hat{x}_{1}\ \hat{x}_{2}]^{T}$, $e=[(\bar{y}-\hat{y})\ (x_{2}-\hat{x}_{2})]^{T}$,  $\hat{x}_{i}\in\mathbb{R}$, and the diagonal matrix $\Gamma=\mbox{diag}(c_{1},c_{2})$, with $c_{1},c_{2}>0$. Finally, the gain $c_{3}$ of the discontinuous term can be chosen as described in \cite{robustobs}. 

An interesting feature of this observer is that, by filtering the discontinuous term, it is possible to identify disturbances/anomalies. This is shown as follows. First, assume that the measurement is free of noise, i.e., $\eta(t)=0$ in (\ref{outputb}), and define the observation errors
\begin{equation}
e_{1}=\bar{y}-\hat{y},\  \  \  \mbox{and}\ \ \ \ e_{2}=x_{2}-\hat{x}_{2}.
\end{equation}
Then, by considering the output vector $C=[1\ 0]$, the corresponding observation error dynamics is described by
\begin{align}
\dot{e}_{1}&=e_{2}-c_{1}e_{1}+\dot{\delta},\label{ob1}\\
\dot{e}_{2}&=-(a+c_{2})e_{1}-be_{2}-c_{3}\mbox{sign}(e_{1})+a\delta.\label{ob2}
\end{align}
When the system reaches the discontinuous surface, it follows that $e_{1}=\dot{e}_{1}=0$. Substituting this into (\ref{ob1}) it follows that on the discontinuous surface, the observation error $e_{2}$ satisfies
\begin{equation}\label{e21}
    e_{2}=-\dot{\delta},
\end{equation}
and consequently, it also holds that 
\begin{equation}\label{e22}
    \dot{e}_{2}=-\ddot{\delta}.
\end{equation}

Next, by using the equivalent control method \cite{utkin}, it follows from (\ref{ob2})-(\ref{e22}) that, on the discontinuous surface, the filtered version of the discontinuous term $c_{3}\mbox{sign}(e_{1})$ satisfies
\begin{equation}
    \overbar{c_{3}\mbox{sign}(e_{1})}=\ddot{\delta}+a\delta+b\dot{\delta},
\end{equation}
where the over bar denotes filtering.

In order to obtain $\overbar{c_{3}\mbox{sign}(e_{1})}$, we use the low-pass Butterworth filter, written in state-space form
\begin{align}
    \dot{x}_{f}&=A_{f}x_{f}+B_{f}c_{3}\mbox{sign}(e_{1}),\label{fil1}\\
    y_{f}&=C_{f}x_{f},\label{fil2}
\end{align}
where $x_{f}=[x_{f1}\ x_{f2}]^{T}$ and
\begin{align}
     A_{f}&=\left[\begin{array}{cc}0&1\\-\omega_{c}^{2}&-1.4142\omega_{c}\end{array}\right],\ \ B_{f}=\left[\begin{array}{c}0\\\omega_{c}^{2}\end{array}\right], \label{fil3} \\ C_{f}&=\left[\begin{array}{cc}1&0\end{array}\right], \label{fil4}
\end{align}
where $\omega_{c}$ is the cut-off frequency of the filter. Then, by choosing a $\omega_{c}$ that minimizes the phase delay, it is possible to assume that
\begin{equation}\label{filto}
    y_{f}\approx\ddot{\delta}+a\delta+b\dot{\delta}.
\end{equation}

\subsection{Anomaly output detector with robust observer and filter}\label{filin}
In order to design the anomaly output  detector, it is worth nothing that the output $y_{f}$ of the filter, see (\ref{filto}), can indeed be seen as a \textit{residual} because it contains information about the  attack/anomaly in the output of the plant.

Then, similar to (\ref{detector}), we define the following detector
\begin{equation} \label{detector2}
\left\{\begin{aligned}
	\left|y_{f}(t)\right| \leq \alpha_{f} &\longrightarrow \text{no alarm} \quad\longrightarrow\quad alarm=0, \\
	\left|y_{f}(t)\right| > \alpha_{f} &\quad\longrightarrow \text{alarm}\quad\longrightarrow\quad alarm=1.
\end{aligned}\right.
\end{equation}
For the time being, we do not have a formal procedure for determining the upper threshold $\alpha_{f}$. Instead, the value of $\alpha_{f}$ is numerically determined  by directly measuring the output of the filter, for sufficiently long time, and then the value of $\alpha_{f}$ will correspond to the $L_{\infty}$-norm computed from the obtained measurement.  
\begin{remark}
Note that for the free-noise case, i.e., $\eta=0$, and considering that the attack is constant, i.e., $\delta:=\delta_{0}$, with $\delta_{0}\in\mathbb{R}$, then the output of the filter, see (\ref{filto}), satisfies
\begin{equation}\label{fre}
   y_{f}\approx a\delta_0.
\end{equation}
Hence, in this case, it is possible  to reconstruct the attack. Note that this fact opens the possibility of using the estimated attack in the control in order to make the system/plant immune to constant attacks by preventing the drift of the estimation error. However, it should be noted that in order to estimate the attack $\delta_{0}$, it is necessary to have knowledge of parameter $a$, which is a parameter of the plant. 
\end{remark}

\section{EXAMPLE} \label{sec:example}
This section presents a numerical example, which illustrates the performance of the detectors presented in Sections \ref{sec:residual_detection} and \ref{sec:discontinuous_observer}. In particular, we consider a second order system of the form (\ref{eq:dLTIsystem}) with 
\begin{equation}\label{22}
A=\left[\begin{array}{cc}0&1\\-4&-20\end{array}\right], \ B=\left[\begin{array}{c}0\\1\end{array}\right],\ C=\left[\begin{array}{cc}1&0\end{array}\right].
\end{equation}
Furthermore it is assumed that the output is influenced by zero-mean Gaussian noise $\eta(t) \sim \mathcal{N}(0,R)$.

\subsection{Residual-Based Methods}

We now demonstrate these tools and provide a comparison between the filtered and unfiltered version of the chi-squared detector. The system matrices are provided above; here we use a control gain $K=[1,1]$ and an observer gain $L=[0,2]'$. The sampling time is $T=0.001$ and the corresponding measurement noise covariance is $\Sigma_\eta=2$.

The idea behind this comparison is that filtration makes the covariance of the filtered residual smaller and removes the high frequency content of the measurement noise, therefore, the attack will be more identifiable when filter is applied. This ease of detecting the attack is apparent when the alarm rate corresponding to the filtered detector deviates more significantly from the desired/tuned false alarm rate, indicating the detector's sensitivity to the attack is higher. We consider a simple attack to demonstrate this point,
\begin{equation}
    \delta(t)=1.
\end{equation}

\begin{figure}[t]
\centering
\includegraphics[width=0.79\linewidth]{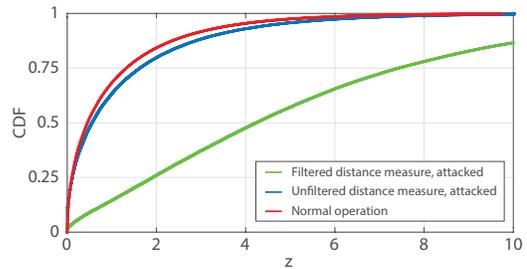}
\caption{This figure shows the performance of the proposed filter-based detector. Using a threshold of $\alpha=3.84$ both detectors have been tuned to provide a $\mathcal{A}^*=0.05$ (5\%) false alarm rate. The green curve plots the cumulative distribution function of the distance measure under attack when computed from the filtered residual $\rho_k$ and exhibits a dramatic difference from what the distribution looks like under normal operation (red). This is quantified by a higher than expected false alarm rate $\mathcal{A}=0.55$ (55\%). In contrast, the blue curve plots the cumulative distribution function of the distance measure under attack when computed from the residual without filtering. Not only does the distribution stay quite similar under attack, but the alarm rate is $7\%$ which is only $2\%$ bigger than the false alarm rate.}\label{fig:cdf}
\end{figure}

\begin{figure}[hb!]
\centering
\includegraphics[width=0.73\linewidth]{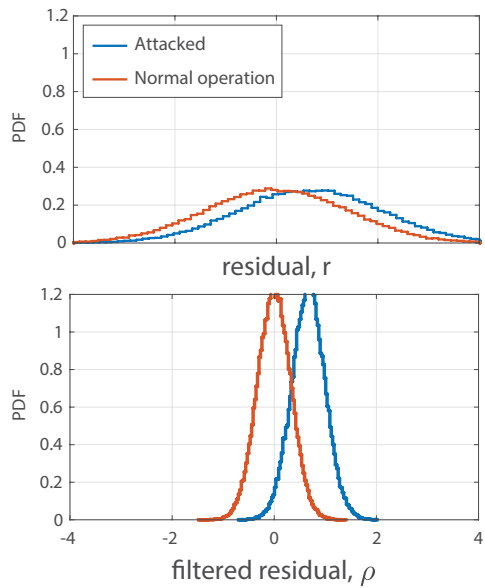}
\caption{The probability density functions, since they are not normalized, aptly demonstrate the role of shrinking covariance in the detection problem. The difference between the attacked and nominal behavior is more dramatic in the filtered residual $\rho_k$ (lower panel) than in the residual $r_k$ (upper panel).}\label{fig:pdf}
\end{figure}

Figure \ref{fig:cdf} shows the cumulative distribution of distance measure calculated from $r_k$ under attack (blue) compared with the cumulative distribution of the distance measure under no attacks (red). Because the distributions are closely similar, the conventional chi-squared detector is not able to easily distinguish the attack. This figure also compares the cumulative distribution of the distance measure calculated from the filtered residual $\rho_k$ (green). In this case the small attack significantly changes the distribution of the distance measure and detection is easy. This is captured quantitatively by comparing the false alarm rates. Here the detectors were tuned to provide $\mathcal{A}^*=0.05$ (5\%) false alarms (i.e., under no attacks) by selecting the threshold as $\alpha=3.84$. 

Under attack the conventional detector's alarm rate was 7\% whereas the filtered detector's alarm rate was 55\%, demonstrating the dramatic difference in sensitivity provided by the additional filtration.

Because the residual and filtered residual are not themselves normalized (as opposed to the distance measure which is normalized), the role that covariance plays is quite clear in Fig. \ref{fig:pdf}. Here the reduced covariance of the filtered residual clearly distinguishes the attack scenario from the normal operation. This distinction is less obvious in the unfiltered residual.

\subsection{Discontinuous Observer}
For the case of the observer-based detector discussed in Section \ref{sec:discontinuous_observer}, we consider observer (\ref{obsi1})-(\ref{obsi2}) with $A$, $B$, and $C$ as given in (\ref{22}) and the gains $c_{1}=c_{2}=5$, and $c_{3}=12$. For this choice of $c_{i}$, $i=1,\ldots,3$, the observation error is globally asymptotically stable for the noise free ($\eta=0$) and attack free ($\delta=0$) case, see \cite{robustobs}.  Furthermore, the chosen cut-off frequency $\omega_{c}$ for the low-pass filter (\ref{fil1})-(\ref{fil4}) is $\omega_{c}=12$ [rad/s].

\begin{figure}[t]
	\centering
		\includegraphics[scale=0.52]{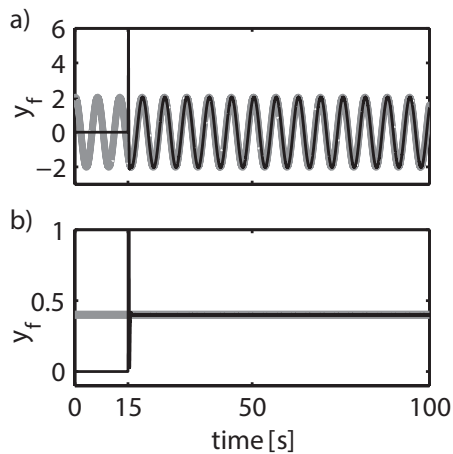}
	\caption{Residual signal for the noise free case. a) Residual for the time varying attack (\ref{varying}). Black line: measured residual. Gray line: predicted residual from (\ref{filto}). b) Residual for the constant attack (\ref{cons}). Black line: measured residual. Gray line: predicted residual from Eq. (\ref{fre}).}
	\label{fig:attacks}
\end{figure}
Next, the performance of the observed-based detector is evaluated for two different attacks, namely a time varying attack and a constant attack. The time varying attack is described by
\begin{equation}\label{varying}
\delta(t)=\left\{\begin{array}{ccc}0&\mbox{if}&t\leq15,\\
                                    0.1\sin{t}&\mbox{otherwise},&\end{array}\right.
\end{equation}
whereas the constant attack is given by
\begin{equation}\label{cons}
\delta(t)=\left\{\begin{array}{ccc}0&\mbox{if}&t\leq15,\\
                                   0.1&\mbox{otherwise}.&\end{array}\right.
\end{equation}
Figure \ref{fig:attacks} shows the numerical and theoretical residuals $y_{f}$. Clearly, in both cases, the measured residual corresponds to the predicted residuals (\ref{filto}) and (\ref{fre}). The numerical results have been obtained  by using Simulink (MATLAB) and the solver Runge-Kutta with fixed step size of 0.001.

On the other hand, for the case of noisy output, it is assumed that the output contains zero-mean Gaussian noise $\eta$ with covariance $\Sigma_\eta=0.001$. In this case, by using the empirical procedure mentioned in Subsection \ref{filin}, it has been obtained that the threshold value of the detector (\ref{detector2}) is $\alpha_{f}=1.55$. It can be shown that if the attacks (\ref{varying}) and (\ref{cons}) are added to the noisy output, then the detector raises an alarm, i.e., those attacks can be identified.

\section{CONCLUSIONS}

In this paper we've introduced two techniques to exploit the a priori knowledge that the noise is a high frequency disturbance to make the attack/anomaly detection task easier. In the first case, we retrofit the conventional chi-squared detector with a filter to reduce the covariance of the detector's distance measure statistic. In the second case, we adopted a robust observer with a discontinuous term that reveals the presence of an attack or anomaly in the system. As fault and anomaly detection tools are poised to bridge to industry, such practical techniques are compelling ways to boost performance and constrain potential attackers even further.

\addtolength{\textheight}{-12cm}   




\bibliographystyle{IEEEtran}
\bibliography{security}

\begin{thebibliography}{10}
\providecommand{\url}[1]{#1}
\csname url@samestyle\endcsname
\providecommand{\newblock}{\relax}
\providecommand{\bibinfo}[2]{#2}
\providecommand{\BIBentrySTDinterwordspacing}{\spaceskip=0pt\relax}
\providecommand{\BIBentryALTinterwordstretchfactor}{4}
\providecommand{\BIBentryALTinterwordspacing}{\spaceskip=\fontdimen2\font plus
\BIBentryALTinterwordstretchfactor\fontdimen3\font minus
  \fontdimen4\font\relax}
\providecommand{\BIBforeignlanguage}[2]{{%
\expandafter\ifx\csname l@#1\endcsname\relax
\typeout{** WARNING: IEEEtran.bst: No hyphenation pattern has been}%
\typeout{** loaded for the language `#1'. Using the pattern for}%
\typeout{** the default language instead.}%
\else
\language=\csname l@#1\endcsname
\fi
#2}}
\providecommand{\BIBdecl}{\relax}
\BIBdecl

\bibitem{Patton_1}
J.~Chen and R.~J. Patton, \emph{Robust Model-based Fault Diagnosis for Dynamic
  Systems}.\hskip 1em plus 0.5em minus 0.4em\relax Norwell, MA, USA: Kluwer
  Academic Publishers, 1999.

\bibitem{Cardenas}
A.~C\'{a}rdenas, S.~Amin, Z.~Lin, Y.~Huang, C.~Huang, and S.~Sastry, ``Attacks
  against process control systems: Risk assessment, detection, and response,''
  in \emph{Proceedings of the 6th ACM Symposium on Information, Computer and
  Communications Security}, 2011, pp. 355--366.

\bibitem{mo2014detecting}
Y.~Mo, R.~Chabukswar, and B.~Sinopoli, ``Detecting integrity attacks on scada
  systems,'' \emph{IEEE Transactions on Control Systems Technology}, vol.~22,
  no.~4, pp. 1396--1407, 2014.

\bibitem{Mo_3}
Y.~Mo and B.~Sinopoli, ``On the performance degradation of cyber-physical
  systems under stealthy integrity attacks,'' \emph{IEEE Transactions on
  Automatic Control}, vol.~61, pp. 2618--2624, 2016.

\bibitem{Guo2016}
Z.~Guo, D.~Shi, K.~H. Johansson, and L.~Shi, ``{Optimal Linear Cyber-Attack on
  Remote State Estimation},'' \emph{IEEE Transactions on Control of Network
  Systems}, vol.~PP, no.~99, pp. 1--10, 2016.

\bibitem{milovsevic2018quantifying}
J.~Milo{\v{s}}evi{\'c}, D.~Umsonst, H.~Sandberg, and K.~H. Johansson,
  ``Quantifying the impact of cyber-attack strategies for control systems
  equipped with an anomaly detector,'' in \emph{2018 European Control
  Conference (ECC)}.\hskip 1em plus 0.5em minus 0.4em\relax IEEE, 2018, pp.
  331--337.

\bibitem{umsonst2018anomaly}
D.~Umsonst and H.~Sandberg, ``Anomaly detector metrics for sensor data attacks
  in control systems,'' in \emph{2018 Annual American Control Conference
  (ACC)}.\hskip 1em plus 0.5em minus 0.4em\relax IEEE, 2018, pp. 153--158.

\bibitem{RuthsACC2018_Hidden}
C.~Murguia and J.~Ruths, ``On reachable sets of hidden cps sensor attacks,'' in
  \emph{2018 Annual American Control Conference (ACC)}.\hskip 1em plus 0.5em
  minus 0.4em\relax IEEE, 2018, pp. 178--184.

\bibitem{RuthsACC2018_Windowed}
R.~Tunga, C.~Murguia, and J.~Ruths, ``Tuning windowed chi-squared detectors for
  sensor attacks,'' in \emph{2018 Annual American Control Conference
  (ACC)}.\hskip 1em plus 0.5em minus 0.4em\relax IEEE, 2018, pp. 1752--1757.

\bibitem{RuthsACC2018_Comparison}
N.~Hashemi, C.~Murguia, and J.~Ruths, ``A comparison of stealthy sensor attacks
  on control systems,'' in \emph{2018 Annual American Control Conference
  (ACC)}.\hskip 1em plus 0.5em minus 0.4em\relax IEEE, 2018, pp. 973--979.

\bibitem{RuthsACC2019_nongaussian}
N.~Hashemi and J.~Ruths, ``Generalized chi-squared detector for lti systems
  with non-gaussian noise,'' in \emph{2019 Annual American Control Conference
  (ACC)}.\hskip 1em plus 0.5em minus 0.4em\relax IEEE, 2019.

\bibitem{robustobs}
\BIBentryALTinterwordspacing
D.~I. Rosas~Almeida, J.~Alvarez, and L.~Fridman, ``Robust observation and
  identification of ndof lagrangian systems,'' \emph{International Journal of
  Robust and Nonlinear Control}, vol.~17, no.~9, pp. 842--861. [Online].
  Available: \url{https://onlinelibrary.wiley.com/doi/abs/10.1002/rnc.1156}
\BIBentrySTDinterwordspacing

\bibitem{Carlos_Justin2}
C.~Murguia and J.~Ruths, ``Cusum and chi-squared attack detection of
  compromised sensors,'' in \emph{proceedings of the IEEE Multi-Conference on
  Systems and Control (MSC)}, 2016.

\bibitem{utkin}
V.~I. Utkin, \emph{Sliding Modes in Control and Optimization}.\hskip 1em plus
  0.5em minus 0.4em\relax Springer-Verlag Berlin Heidelberg, 1992.

\end{thebibliography}

\end{document}